# BEACON-BASED UPLINK TRANSMISSION FOR LORAWAN DIRECT TO LEO SATELLITE INTERNET OF THINGS

Mohammad Al mojamed

Computer Science Department, College of Engineering
and Computing, Al-Qunfudhah, UMM Al-QURA University, Saudi Arabia

## ABSTRACT

*Direct-to-satellite IoT (DtS-IoT) communication structure is a promising solution to provide connectivity and extend the coverage of traditional low-power and long-range technologies, especially for isolated and remote areas where deploying traditional infrastructure is impracticable. Despite their bounded visibility, the Low Earth Orbit (LEO) satellites complement the terrestrial networks, offering broader gateway coverage and terrestrial network traffic offloading. However, the dynamics of LEO and the nature of such integration come with several challenges affecting the efficacy of the network. Therefore, this paper proposes Beacon-based Uplink LoRaWAN (BU-LoRaWAN) to enhance satellite-terrestrial communication efficiency. The proposed scheme exploits the LoRaWAN class B synchronization mechanism to provide efficient uplink transmission from LoRaWAN devices placed on the ground to satellite gateways. BU-LoRaWAN proposes an uplink transmission slot approach to synchronize ground devices' uplink traffic with LEO-based orbiting gateways. It also uses a queue data structure to buffer end devices' ready-to-send packets until the appropriate moment. BU-LoRaWAN avoids possible transmission collision by optimizing a random transmission slot for an end device within the beacon window. The proposed system is implemented and evaluated using OMNeT++ network simulator and FLoRaSat framework. The result demonstrates the feasibility of the proposed system. BU-LoRaWAN achieves better performance compared to the standard LoRaWAN, which manages to deliver almost double the traffic delivered by the standard one.*

## 1. INTRODUCTION

Applications of Internet of Things (IoT) are bringing essential changes in our lives, improving operational awareness and real-time decisions. A large variety of application scenarios do exist, leading to diverse communication technology requirements. IoTs applications include meteorological monitoring, geological and environmental phenomena monitoring, animal tracking, remote wildlife monitoring, oil and gas monitoring, and transportation monitoring. However, various IoT scenarios require the deployment of long-range and low-power IoT devices in widespread and remote areas such as Oceans, Seas, the Arctic, the Antarctic, and large forests. Therefore, to cover such large and remote communication areas, the satellite technology approach can be applied while retaining the features of long-range and low-power devices on the ground.

Satellites can facilitate ubiquitous coverage, leading to revolutionary IoT applications. This is driven by the affordability and availability of the Low Earth Orbit (LEO) satellites. LEO satellites are more appropriate than the Geostationary Earth Orbit (GEO) satellites for IoT applications due to expensive deployments and GEO's high latency resulting from the long distance. A shorter delay is being offered by LEO due to orbiting closer to the Earth compared to GEO satellites. Undoubtedly, LEO satellites have been recently recognized as a data collection and communication provider tool for truly worldwide and cutting-edge IoT solutions [1], [2]. However, their inherent feature of orbiting closer to the Earth results in bounded visibility [3]. Consequently, large LEO constellations are required to overcome this issue and provide better coverage.

LEO satellites play a vital role as a key enabler for IoT connectivity to extend the coverage of traditional low-power and long-range technologies [4]. This can be achieved by connecting constrained ground-based devices to orbiting gateways, establishing what is known as direct-to-satellite IoT (DtS-IoT) communication [5]. In the context of the Low-Power Wide Area Network (LPWAN), the Long-Range Wide Area Network (LoRaWAN) is widely adopted due to its advantageous characteristics, such as long-distance transmission and low power consumption [6]. LoRaWAN is a star-of-star topology that offers single-hop communication between edge devices and gateways. LoRa is the used physical layer for LoRaWAN, which is based on the Chirp Spread Spectrum (CSS) to allow the transmission to large distances with interference, doppler effects, and fading resistance [7][8][9]. The feasibility of LoRaWAN long-distance communication has been proven, and several experiments have demonstrated the ability of LoRaWAN devices to send their traffic up to thirteen hundred Kilometres [10] [11].

Deployed LoRaWAN networks in remote and harsh environments can exploit the integration of terrestrial and satellite networks to overcome local network disconnectivity issues. The isolation of the network may occur due to different causes, such as when the network itself is installed in a remote area or when the communication infrastructure is damaged due to a natural disaster. Given that only 10% of the earth's surface is covered by terrestrial connectivity due to different limitations [12], the integration of these technologies can fill the connectivity gap where no ground infrastructure is available. The integration of terrestrial-satellite networks has emerged as a promising solution for backhauling LoRaWAN networks. LEO satellites, thus, complement the terrestrial networks, offering wider gateway coverage and terrestrial network traffic offloading [13], [14]. However, integrating LEO satellites and terrestrial networks brings several challenges and requires a smooth design for an efficient solution.

The contribution of this paper is to address the efficiency of LoRaWAN integrated network with the LEO satellite. The presented study proposes a Beacon-based Uplink LoRaWAN (BU-LoRaWAN) that enhances the reliability of satellite-terrestrial communication. The novelty of the system lies in the fact that the proposed system utilizes modified LoRaWAN class B functions to serve a different goal that the class was not developed to achieve initially. The beaconing technique of Class B is developed to allow the network server to know when to send downlink traffic to the end device. However, the proposed system utilizes this to know when to transmit uplink traffic at no extra cost in a LoRaWAN-LEO integrated network. Hence, utilizing Class B beaconing technique for uplink transmission in the scarce network.

Furthermore, the developed system identifies some parts of the beacon window as possibly an uplink transmission window and proposes uplink slots within the window. It also introduces randomization during the scheduling of uplink traffic slots within each window to avoid collisions. The BU-LoRaWAN scheme finds the time when the satellite gateways are reachable, i.e., when they fly over current LoRaWAN devices. Thus, end devices can utilize this opportunity to transmit their uplink traffic. The proposed system is evaluated using network simulation tools. Results indicate that better performance is achieved compared to standard LoRaWAN operating class B mode.

The rest of the paper is structured as follows. Section 2 introduces the related studies that address the use of LEO and terrestrial LoRaWAN networks. The network structure of LEO and terrestrial network integration is then discussed in section 3. Section 4 introduces the proposed system. The proposed system evaluation is presented in section 5. First, the simulation setup and the scenario configuration are provided, and then the result is introduced and discussed. Finally, section 6 concludes the paper and highlights possible future work.

## 2. RELATED WORKS

Several studies have been conducted addressing the integration of LoRaWAN and LEO communications. Several issues were targeted, including the adopted LoRa signal modulation, LEO constellation design, shared medium access, and transmission scheduling. Moreover, some studies have evaluated the general ground-to-satellite communication architecture or tested some specific scenarios' feasibility. The following introduces and highlights these related studies.

The challenge of sparse direct LoRaWAN to satellite communication was addressed in [15] by introducing a transmission probability function to control the cardinality of devices set within a frame. The proposed solution is based on framed slotted Aloha as the medium access technique. It assumes that satellites are aware of prior information of contending LoRaWAN devices regarding their positions and transmission needs. Thus, gateways are fully aware of the number of contending devices within each frame. The cardinality is then used by the gateway satellite to calculate the transmission probability function and schedule a slot for each ground device. The transmission probability is then attached to satellite beacon messages and broadcast to end devices.

Satellite trajectory knowledge has been exploited in [16] to enhance the performance of direct-to-satellites IoTs. Multiple uplink transmission policies were proposed to work on legacy LoRa and Long-Range Frequency Hoping Spread Spectrum (LR-FHSS) utilizing the satellite's location information. All the proposed policies assume that end devices have the ability by some means to derive the current distance between them and the passing by LEO gateways as well as the estimated distance in the immediate future. Different ways can be used or combined to infer the distance, including RSSI extrapolation, determining the speed via processing measured frequency shift, and exploiting orbital parameters to get an accurate gateway trajectory. The proposal then assumed the presence of a central scheduler to compute the transmission schedule for each of the end devices. The central scheduler is suggested to be on the satellites or the network server and is aware of all the system states, such as the pattern of traffic and packet sizes.

One of the main issues in the direct-to-satellite communication paradigm is the availability of gateways to ground-based end devices. The work in [17] has proposed an algorithm to address the discontinuous coverage availability issue using the well-known Time Division Multiple Access (TDMA) to access the shared medium following LoRaWAN class A mode. Participating devices can connect to the gateway and transmit their traffic according to a predefined schedule. The scheduling is based on different factors, including the availability of satellites and specific visiting times for a particular device. Two different approaches were used to decide on a transmission slot for a device: First Come, First Served, and Fair policy. However, the process of calculating and deciding the scheduling is under the responsibility of the Network server rather than the end device and the gateways. The proposal assumed that not only the location of each end device is priorly known to the network server but also the visibility time for each device is known. Moreover, the algorithm assumes excluding the way of exchanging the scheduling table and the configuration parameters between the network server and end devices through the satellite gateway.

A MAC joint optimization scheme was proposed in [18] for indirect LoRaWAN to satellite communication. Gateways are located in the ground and allocated a time slot by the satellite, which acts as the central unit in the network. A ground LoRaWAN gateway uses a continuous forward time slots allocation scheme to allocate time slots for its sensor nodes within its coverage based on the slots given by the satellite. The proposal considers a geographic priority allocation scheme to utilize satellite resources. According to the employed geographic priority allocation scheme, a gateway is required to send its geographic location parameters to the satellites. The satellite then allocates the gateway a fixed slot, which will be used as the basis for the gateway for servicing its sensor nodes. The proposal was simulated and showed its ability to reduce the delay and improve network throughput.

The work in [19] investigates the effect of satellite inclination angles at different satellite altitudes to find the most suitable design to provide the best possible IoT coverage. The study used LoRa as the basis for an IoT terminal connected to a space satellite. The study used a network simulator and paid most of the attention to the effect of inclination and altitude on direct-to-satellite communication. It was found that with low inclination, the number of satellites can be reduced while maintaining good coverage. The work in [2] developed a dynamic parameters reconstitution method to enhance ground LoRaWAN to satellite communication. The methodology involved several calculation processes, including satellite-to-earth elevation angle calculation, satellite-to-earth distance calculation, and satellite-to-earth link budget calculation. The calculation of the elevation angle uses LEO satellite orbit information and the geographic location of the ground terminals. The proposal did not state how such information is obtained. The methodology was evaluated using a network simulator, and the proposal was shown to be able to overcome the issue of satellite motion in ground-to-satellite communication.

The required in-orbit infrastructure has been investigated in [20]. The study explored the possibility of reducing the number of in orbit satellites while maintaining massive ground-to-satellite communications on regional and global levels. It exploited some of the features of LoRaWAN, such as beacon less and power saving mode to implement a sparse constellation design using less than a quarter of required in-orbit satellites in traditional dense constellations. The design is based on the delay tolerance nature of the IoT to satellite communication systems. Accordingly, LoRaWAN devices were left out of coverage for up to two hours without detaching devices from the network. Class B mode was used and exploited to let the devices know the availability of passing by satellites.

The performance of LoRa –CSS was investigated by the Space–D project conducted by the Dubai Electricity and Water Authority (DEWA). DEWASAT is the name of the launched CubeSat. The work in [21] is one of a series of published works [22] by the group and mainly focuses on evaluating the link margin of LoRa CSS modulation. One terminal and one CubeSat, DEWASAT-1, were used to assess the connection possibility using different parameters, including spreading factors, antenna gain, and coding rate. The study concluded that uplink communication between the ground terminal and the satellite is achieved. The work in [23] investigated the feasibility of direct-to-satellite communication for real-life scenarios, such As maritime autonomous surface ships, using LoRaWAN. The scenario is designed based on actual data from the Arctic Data Centre, including the number of ships, their distribution, and the pattern of ship traffic. The study investigates using two LoRaWAN modulation schemes: Chirp Spread Spectrum and long-range frequency hopping spread spectrum with different configuration parameters. It also studied the performance of such scenarios using multiple satellites. It concluded that direct-to-satellite communication is achievable using both modulation schemes. However, the use of long-range frequency hopping spread spectrum gives better success probability compared to the conventional LoRa Chirp Spread Spectrum modulation.

The potential feasibility of LR-FHSS LoRaWAN modulation was also investigated in [12]. The focus of the paper was to examine the LR-FHSS for direct-to-satellite networks. The study started by proposing an analytical model that describes the performance of LR-FHSS, assuming that gateways are located on LEO satellites with characteristics like the Iridium satellites. The study then proposed a simulation model using MATLAB-based Monte Carlo simulation to validate the analytical model and allow complex scenarios. The work concluded that the performance of such direct-to-satellite communication can be enhanced when the capture effect is considered as packets are lost due to losing the headers.

Modifications of the LoRa physical layer were proposed in [24] to enhance the demodulation of the received LoRa signal. A differential chirp spread spectrum (DCSS) was proposed to alleviate the issue of decoding sensitivity to frequency and time synchronization error in the LoRa signal. Thus, the proposed technique entitles the receiver to tackle ultra narrowband signal since no restriction on maximum frequency offset is introduced. The technique also achieved better performance compared to CSS despite the presence of a Doppler shift, which makes it a good choice for LoRa to use in satellite communication. Another work that addressed the integration of LoRaWAN with satellite through the modification of the LoRa physical layer signal is [25]. A differential chirp spread spectrum was proposed to better suit satellite communication as it is insensitive to carrier frequency offset caused by Doppler effects. The proposed receiver uses a time synchronization approach to introduce no limitation on the carrier frequency offset and decode the signal accurately. The proposed differential chirp spread spectrum eliminates the use of the preamble and the start of frame delimiter to estimate the offset. Instead, it uses the N up-chirps of the preamble to detect the signal and estimate the fractional offset.

In order to avoid the Doppler frequency shift in IoT satellite communication, the work in [26] has also proposed a differential modulation scheme. A Maximum likelihood Sequence Detection (MLSD) Demodulation Scheme is further developed for the satellite receiver to overcome the error spread phenomenon resulting from the proposed differential modulation algorithm. Through simulation, the work concluded that the proposed design gives better performance than the standard LorRa modulation scheme. The possibility of the reception of LoRa signal from LEO satellites is studied in [27]. The work also identified LoRa signal properties that can be exploited for synchronization purposes. The work designed a satellite-based receiver architecture to handle the impairment of satellite signal detection, such as Doppler rate and Doppler shift. It also uses interference cancellation mechanisms to improve the satellite's ability to receive LoRa signal.

LoRaWAN-CubeSat signal propagation was investigated and assessed as well in [28]. The study incorporates some of the LoRaWAN features, such as Adaptive Data Rate (ADR), to examine the capability of dynamic signal selection for different scenario conditions. Experiments were conducted to find out the possibility of receiving LoRaWAN traffic from long distances. Attenuators were used to simulate the effect of path loss. The study concluded that LoRaWAN gateways can be used on CubeSat and could receive uplink traffic from ground terminals. However, the study admitted that the adoption of LoRaWAN for ground-to-satellite connection should consider several fields and parameters that may influence the connection.

LORSAT was also developed in [29] as an emulation testbed. It combined LoRaWAN network real devices with emulated satellites to implement entire LoRaWAN – Satellites networks. It considered real environmental factors and actual conditions to validate such a design. The work concluded the practical feasibility of the developed design. Moreover, the Massive Machine-type connectivity (mMTC) to LEO was suggested by [30-31] for offshore wind farm monitoring. More on mMTC can be found in [40]. The architecture is based on LoRaWAN, where sensors are placed on the wind turbines and Gateways on the satellites. The designed architecture was evaluated using a simulator. Real-life wind turbine positions and traffic patterns were used to simulate the most challenging wind farm near Denmark's shore. The work concluded that the feasibility of such architecture is relatively low due to the high number of collisions.

The existing works have addressed several challenges related to the integration of LEO and LoRaWAN, one of which is transmission scheduling. Different assumptions were made to enable such proposed solutions. This includes gateways' awareness of existing ground contending devices to calculate some proposed criterion, the prior awareness of satellite trajectory, ground devices' awareness of distance to the flying satellites, or the use of a central scheduler. Therefore, there is a need for a distributed, lightweight, and automated transmission scheduling mechanism to enhance the performance of such integration.

## 3. NETWORK ARCHITECTURE

LoRaWAN-based IoT devices can be integrated with LEO satellites by following two possible approaches: direct-to-satellite IoT and indirect-to-satellite IoT [32]. The latter approach involves the use of ground gateways, which will be responsible for ground-to-satellite communication. Thus, the ground network would maintain the same characteristics as LoRaWAN network since devices can send their traffic to the ground-based gateway as in the standard LoRaWAN network. However, the gateway then requires the use of space-domain communication protocols to communicate with satellites. On the other hand, the direct-to-satellite approach places gateways on the orbiting satellites as on-board gateways. Ground IoT devices, therefore, should direct their traffic to the orbiting gateways. Such network architecture makes it more applicable for applications that operate in less accessible areas.

The direct-to-satellite architecture is based on the LEO satellites, which have recently been recognized as a data collection and communication provider tool for truly worldwide and cutting-edge IoT solutions. LEO satellites, flying at altitudes less than one thousand km, are the most appealing satellites for LoRaWAN-based ground networks. An IoT device fitted with a low-cost antenna can communicate with such a satellite. The feasibility of such architecture has been proved by several already deployed satellites, such as LacunaSat [33] and ThingSat [34].

The direct-to-satellite architecture implies that gateways are located on the board of LEO satellites, flying at high speed. Therefore, ground-to-space channel conditions vary excessively, introducing a significant challenge to integrating LoRaWAN with LEO satellites. It is due to the motion of the orbiting satellites. LEO satellites, flying at less than 2000 km altitude, require an orbital period that ranges from 90 minutes to 120 minutes [35]. In the case of flying at less than 1000 km, a satellite will be visible for a specific geographic region in 10 minutes to 3 minutes, depending on the location of the satellite, whether it is a pass over the horizon or a zenithal pass. Thus, the channel conditions between ground devices and the gateways are expected to change dramatically.

Moreover, the coverage area of a moving satellite changes over time, resulting in different served ground devices. To overcome this challenge, satellites are arranged in a constellation, such as the Kuiper project [36] by Amazon and the Starlink project [37]. This would give better coverage for a specific area as one satellite disappears on the horizon, and another one would appear to serve the area. Satellites within a constellation can communicate with each other using what is known as Inter Satellite links (ISL) to relay received data to the core network server at the ground. Fig.1 illustrates the overall network architecture used in this research.

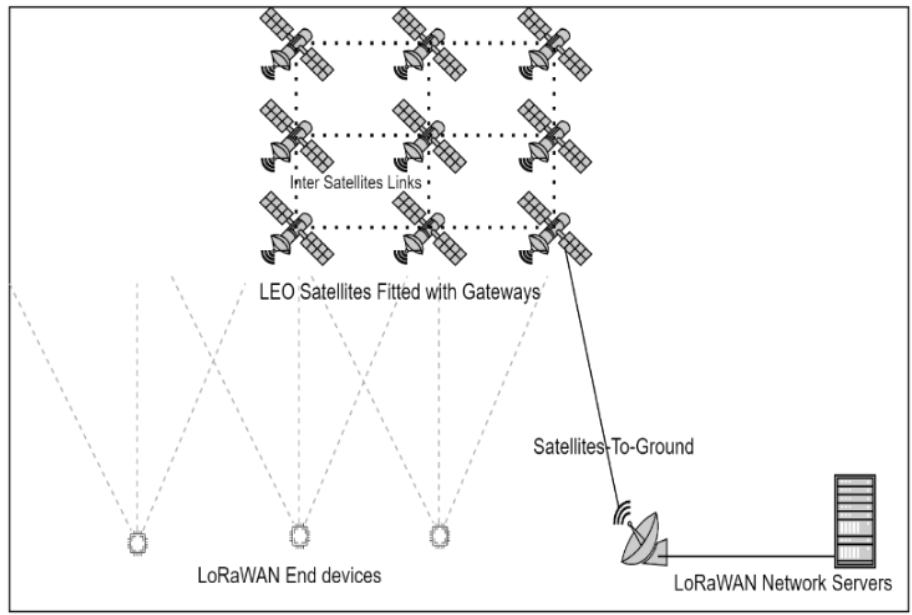

Figure 1: Direct to Satellite LoRaWAN-LEO Architecture

## 4. PROPOSED SYSTEM

The objective of the proposed system, BU-LoRaWAN, is to provide efficient IoTs-LEO communication in environments that offer partial connectivity through the deployment of a constellation with a smaller number of satellites. Such scenarios would require LoRaWAN devices to postpone uplink traffic until the moment when the constellation of LEO satellite passes over the devices' territorial area. The proposed system is based on LoRaWAN class B mode. It utilizes the beaconing scheme that standard class B LoRaWAN uses for downlink communication synchronization. It implements a pull-based traffic transmission as an opportunistic networking technique based on the class B operation. Hence, constellations with few satellites can suffice the operation of the LEO-Terrestrial network. The proposed system provides an uplink scheduling scheme that enhances the performance of LoRaWAN for IoTs LEO communications without introducing considerable modifications to the standard LoRaWAN.

In LoRaWAN class B mode, LoRaWAN devices and gateways are synchronized based on beacon signals to have the end device available for communication at a predictable time. Hence, Network servers will be aware of when to send the downlink traffic to end devices. To achieve such synchronization, gateways broadcast beacons frequently to synchronize possible end devices within the network. The interval between two beacons is known as the beacon period, which is usually 128 seconds. Fig.2 displays the beacon timing for class B mode LoRaWAN. Each beacon period is divided into several slots. 2.12 s, 3 s, and 122.8 s are allocated for the beacon reserved, the beacon guard, and the beacon window, respectively. The beacon window is further divided into small slots of 30 ms, known as ping slots, in which an end device opens its reception window. The beacon guard is used to precede a beacon reserved interval to avoid possible collision between any ping slot traffic and a beacon signal.

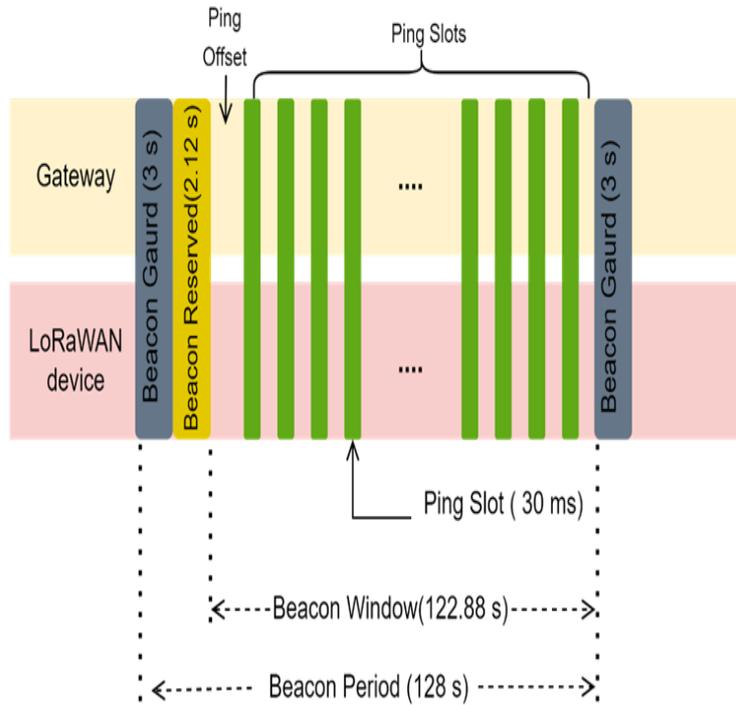

Figure 2: LoRaWAN Class B Beacon Timing

The main issue of deploying LoRaWAN using a Satellite constellation is that an end device may send its uplink traffic at a time when the whole constellation is out of coverage. The proposed scheme aims to find the time when the satellite gateways are reachable, i.e., flying over current LoRaWAN devices. Thus, end devices can utilize this opportunity to transmit uplink traffic. The proposed scheme enhances uplink performance from LoRaWAN devices placed on the ground to the satellite gateways using a class B beaconing scheme.

BU-LoRaWAN implies that an end device is allowed to transmit uplink traffic, providing that it has received a recent beacon signal from one of the constellation gateways. Receiving a beacon indicates that the current ground device is under the coverage of at least one of the constellation gateways. Therefore, an end device in the proposed system defers any request for transmission until a beacon has been received to guarantee the coverage of the gateway. This is done at the MAC layer, where a queue is implemented to enqueue any uplink transmission until the appropriate time.

The proposed uplink transmission window is described in Fig.3, which illustrates the modified beacon timing. The lifecycle of the beacon begins with a 3 s interval as a beacon guard, and the beacon signal should be received during the succeeding 2.12 s beacon reserved period. Following that, the beacon window (122.88 s) starts and ends at the beginning of another beacon guard slot. The beacon window, excluding the Ping Offset interval, is identified as an opportunity time frame for uplink transmission in the proposed scheme. The window is crisscrossed in the figure and labeled with a possible uplink transmission window.

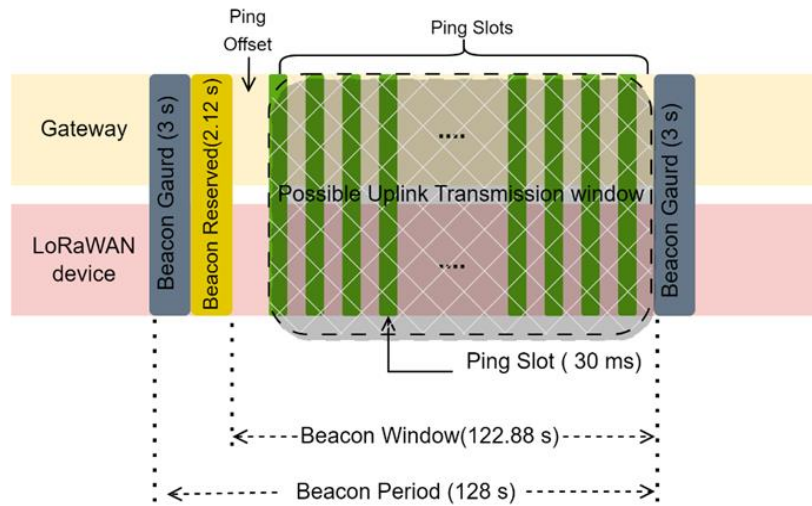

Figure 3: Beacon Timing shows possible uplink transmission time for the proposed system

To avoid uplink transmission collisions, the proposed system picks a random transmission time as an uplink transmission slot based on the summation of two numbers, as shown in Equation 1, in addition to the current time. The first, pingOffset, is the pseudo-random offset enforced at the beginning of a beacon window by LoRaWAN specification to align reception slots. The second operand is a random number calculated by each device. It is randomly chosen where 0 is the lower limit, and the remaining beacon window frame is the upper limit. When the randomly selected time is met, the TX slot begins, and the transmission is initiated.

$$TX\ slot = current\ Time + pingOffset + uniform(0, leftTimeOfCurrentBeaconWindow) \quad (1)$$

The developed scheme utilizes a queue data structure to enqueue any arriving packets at the MAC layer to be transmitted toward one of the available gateways in the constellation. The arrived packets from the higher layer are queued in the queue until the transmission signal is triggered. The queue is inspected frequently based on the occurrence of the internal transmission signal. The transmission timer is calculated on two different occasions. It is mainly calculated once an end device has received a beacon signal. The reception of such a beacon is a sign of being under one of the gateway coverages. Hence, it is an opportunity for delivering the uplink packet. The second occasion for calculating the timer is when the current timer has ended. The new timer is computed within the current beacon window interval whether the queue is empty or not. Consequently, the scheme ensures that it frequently monitors the queue for any recently arrived packets or any upcoming traffic. The flow diagram and the pseudocode are shown in Fig.4 and Table 1, respectively, describing the BU-LoRaWAN procedure for establishing the internal transmission timer.

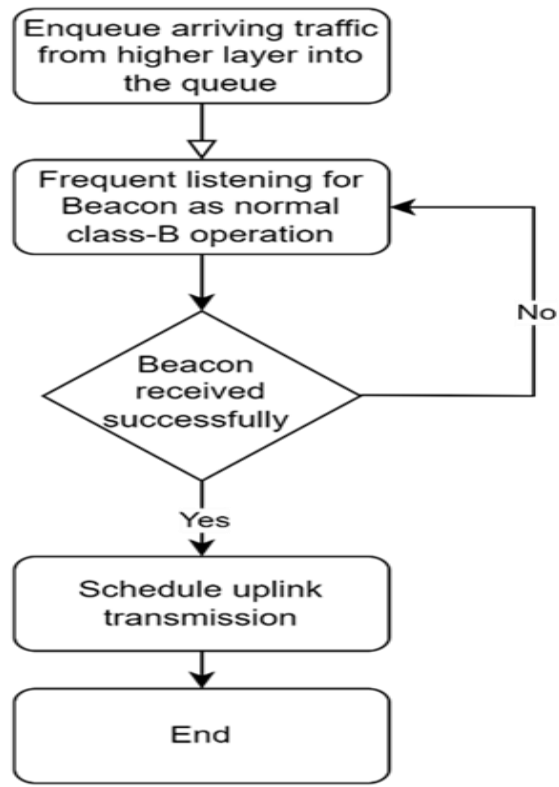

Figure 4: Flowchart of BU-LoRaWAN System

As the network architecture involves sky-crossing constellation, ground LoRaWAN end devices are susceptible to being out of gateway coverage. The proposed scheme handles that by prohibiting transmission if no beacon window is in progress. Instead, it enforces end devices to temporarily store and delay transmission of uplink traffic until chances of being in connection with the LEO constellation arise.

Table 1: Pseudocode for uplink scheduling in the proposed system

| **Algorithm 1: Pseudocode for uplink scheduling in the proposed system** |
|---|
| 1:  PKT: Packet received from higher layer for transmission |
| 2:  Beacon_Rcvd_Flag: Beacon Received Flag |
| 3:  enqueue PKT |
| 4:  Wait for beacon |
| 5:  if Beacon is received successfully |
| 6:      Set Beacon_Rcvd_Flag to 1 |
| 7:      Calculate TX slot based on Eq.1 |
| 8:      Schedule next TX internal signal |
| 9:  else |
| 10:     Set Beacon_Rcvd_Flag to 0 |
| 11: end if |
| 12: if   TX internal signal is fired |
| 13:     Calculate following TX internal signal based on Eq.1 |
| 14:     If queue is not empty and Beacon_Rcvd_Flag is 1 |
| 15:         Pop queue |
| 16:         Initiate PKT transmission |
| 17:     end if |
| 18: end if |

## 5. EVALUATION

### 5.1. Simulation Setup

A series of simulations have been carried out to validate the feasibility and evaluate the performance of the proposed system. The FLoRaSat, the Framework for Lo-Ra-based Satellite, is used [38]. It is an OMNeT++ simulator based on the LoRa and LoRaWAN implementations for space communications. FLoRaSat provides orbital dynamic features that propagate satellites across realistic trajectories. The framework involves an inter-satellites communication model to offer LEO to LEO routing since the considered constellation consists of 16 LEO satellites, each fitted with a LoRaWAN gateway. Inter-satellite communication is used to direct traffic toward one of the gateways that is connected to the ground station. Regarding the channel model, a model is implemented that considers interference, attenuation, and capture effects.

The simulated scenario is similar to the presented scenario in the FLoRaSat paper [39]. It comprises four orbital planes with ascending nodes in 310°-330°-350°-370°, each having four LEO satellites equally separated from each other. The altitude of the constellation is 600 km with $98°$ inclination. LoRaWAN devices are distributed uniformly within an area of a 2000 km radius. The free space path loss model is used as the propagation model to account for signal power reduction as it travels through space. The Spreading Factor SF 12 is used with 125kHz channel bandwidth and four coding rates. Each device is configured to send a 20-byte payload packet every 8-12 minutes. Each single run is repeated ten times, and the plotted result is the derived average of these repetitions. Table 2 shows the simulation parameters.

Table 2: Simulation Parameters

| Parameter | Value |
|---|---|
| LEO satellites | 16 |
| Orbital planes | 4 |
| Satellites per plane | 4 |
| Inclination | $98°$ |
| Satellites Altitude | 600 km |
| Simulation area | Circular with 2000 km radius |
| Number of end devices | 100, 200, 300, 400, 500 |
| Number of gateways | 16 |
| Simulation time [Seconds] | 1200, 2400, 3600, 4800, 6000, 7200 |
| Repetition | 10 |
| Path loss model | Free Space |
| Bandwidth [kHz] | 125 |
| SF | 12 |
| Coding Rate | 4/8 |
| Carrier Frequency | 868 MHz |
| Transmission Power [dBm] | 14 |
| Send Interval [Second] | 8-12 minutes |
| LoRaWAN device class | Class B |
| Packet Size [Byte] | 20 |

## 5.2. Results and Discussion

Fig.5 and Fig.6 show the overall delivery rate as a function of simulation time and network size, respectively. Both figures show that the proposed scheme enhances the efficiency of deploying LoRaWAN with LEO. Fig.5 depicts the ability of the BU-LoRaWAN procedure to maintain a good performance across different simulation periods. The acceptable performance is achieved even though end devices are frequently out of the constellation coverage due to constellation orbit motion. On the other hand, Fig.6 reflects how the standard LoRaWAN is affected due to the disappearance of the satellite fleet. As the simulation period gets longer, the satellite fleet spends more time away from the devices' region, causing coverage disruption. The performance deterioration is due to the continuous transmission of uplink traffic from edge devices while gateways are fading from sight. This is due to the lack of awareness of the availability of moving gateways. Thus, end devices keep transmitting despite being away from the satellites.

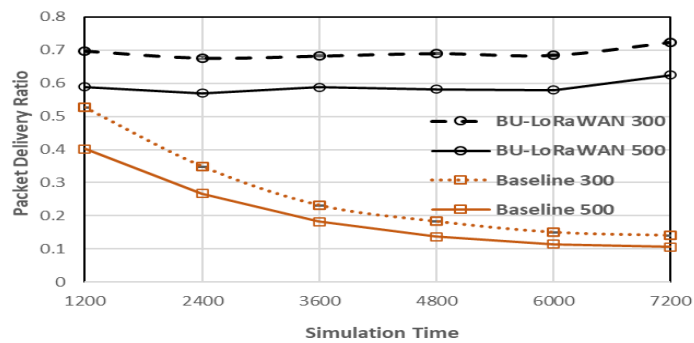

Figure 5: Delivery ratio versus Operation time for 300 and 500 network sizes

The extraction rate of BU-LoRaWAN is also investigated as a function of network size. The results are presented in Fig.6. Both policies, BU-LoRaWAN and LoRaWAN class B baseline, suffer a decrement in delivery rate as the network size increases. Notably, the proposed procedure manages to achieve better performance where around 90% of the traffic is extracted successfully with 100 devices network compared to 80% achieved by the baseline. The decline in the baseline performance can be explained by the occurrence of packet collisions, which is the case with the proposed uplink strategy. Moreover, the transmission of uplink traffic in the absence of an orbiting constellation is observed to be another reason for such worse performance.

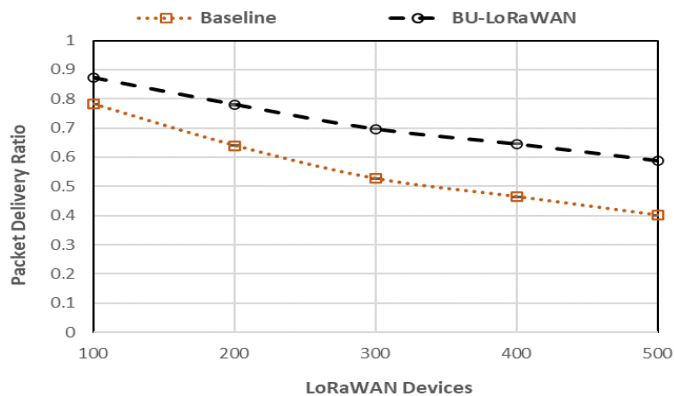

Figure 6: Delivery ratio versus network size for proposed and baseline systems

The number of collisions is depicted in Fig.7 for networks operating for 6000 s. As expected, collisions increase for both systems as the number of participating devices increases. However, fewer collisions occurred in the proposed system for most of the simulated cases while achieving a better extraction rate. This can be justified by the use of the traffic buffering technique in addition to the utilization of most of the beacon window frames for uplink transmission. Therefore, better performance is achieved where some collisions are avoided compared to the baseline.

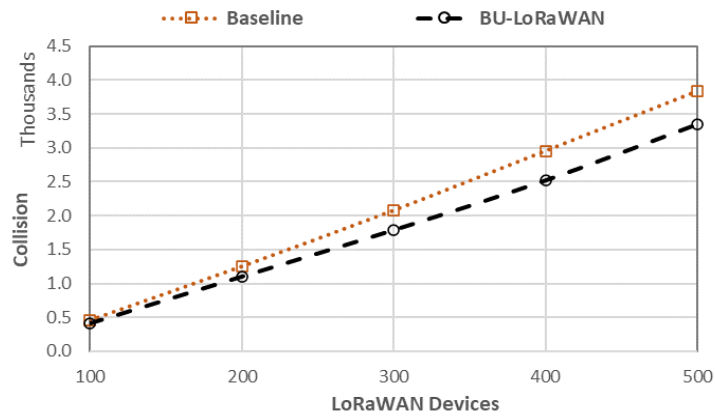

Figure 7: Network collisions versus network size for proposed and baseline systems

Fig.8 shows the number of collisions for 400 network size as a function of different operation periods. Similar collision incidents are observed by the standards LoRaWAN across different simulation times. This is an indication of traffic being transmitted while gateways are out of sight since no increments in collisions are recorded nor an increase in extraction rate is observed. Furthermore, this is justified by the deterioration of the delivery ratio, as shown in previous figures. Contrariwise, the proposed system keeps delivering traffic to the orbiting gateways; hence, chances of collision increment do exist. This is shown in Fig.8, where the proposed BU-LoRaWAN collisions increase for network operating for 7200 s. Additionally, the achieved delivery ratio, Fig.5, indicates that more traffic is arriving at gateways in the proposed one, causing more collisions. However, the obtained better performance comes at the expense of higher collisions.

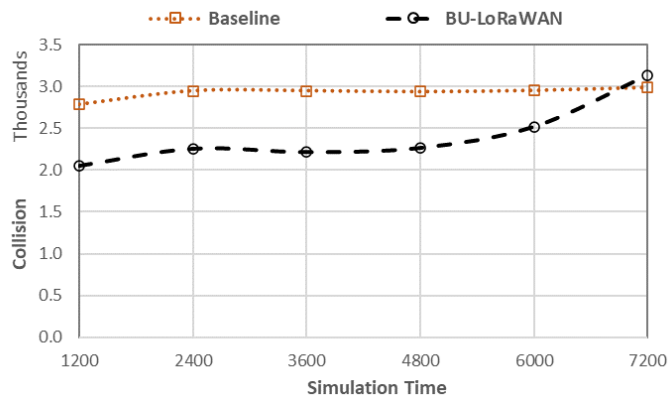

Figure 8: Network collisions versus operation time for 400 network size

## 6. CONCLUSION

The work in this article investigated direct ground-to-satellite communication to enhance the efficiency of LoRaWAN-LEO integrated architecture and provide competent worldwide connectivity for IoT solutions. Thus, this article proposed Beacon-based Uplink LoRaWAN (BU-LoRaWAN) to boost the capability of the ground IoT devices to deliver their traffic to the main application servers through LEO. The proposed system utilizes the beaconing strategy employed by standard LoRaWAN class B for downlink communication synchronization without introducing considerable modifications to the standards. The advantage of the beaconing scheme is, therefore, taken to provide efficient uplink transmission synchronization for edge devices. The uplink transmission in the proposed system is bound to the reception of beacon messages from an orbiting satellite, hence ensuring the device is covered by the satellites. This was achieved by implementing a queue data structure at the MAC layer for LoRaWAN end device. Moreover, the proposed system introduces the use of an uplink transmission slot to allow end devices to transmit their traffic. The uplink transmission slots are calculated within each beacon window randomly to avoid collisions with other devices' traffic. The results of the simulation show the potential feasibility of the proposed system. It outperforms the conventional LoRaWAN due to the employed synchronization of the uplink traffic. The results demonstrate the capability of the proposed system to boost the packet delivery ratio. The network scalability of LoRaWAN-LEO could be further investigated using LR-FHSS modulation, as it provides an additional boost to expand network scalability. A possible future work is also to investigate the calculation mechanism of the random transmission slot further and study its potential effect on collisions that occur.

## CONFLICT OF INTEREST

The author declares no conflict of interest.

**AUTHOR**

**Mohammad Al mojamed** is an associate professor at Computer Science department, College of Engineering and Computing, UMM Al-QURA University. He obtained his PhD from Stirling University, UK in 2016 in Computer Science. His research interests include ad hoc Network, wireless sensor networks, mobility, low power wide area Network, LEO communication, and P2P networks. He can be contacted at mmmojamed@uqu.edu.sa.

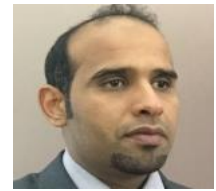